\documentclass[12pt]{iopart}
\usepackage{graphicx}
\usepackage{iopams}

\begin{document}
\title{Analysis of light-induced frequency shifts in the photoassociation of ultracold metastable helium atoms}
\bibliographystyle{unsrt}
\author{M~Portier$^{1,3}$, S~Moal$^{1,3}$, J~Kim$^{1,3}$, M~Leduc$^{1,3}$, C~Cohen-Tannoudji$^{1,3}$ and O~Dulieu$^{2,3}$}
\address{$^1$ Laboratoire Kastler Brossel, CNRS, Ecole Normale Sup\'erieure, Universit\'e Pierre et Marie Curie and Coll\`ege de France,~24 rue Lhomond 75231 Paris Cedex 05, France}
\address{$^2$ Laboratoire Aim\'e Cotton, CNRS, B\^at. 505, Universit\'e Paris Sud, 91405 Orsay Cedex, France}
\address{$^3$ Institut Francilien de Recherche sur les Atomes Froids,~24 rue Lhomond 75231 Paris Cedex 05, France}
\ead{mportier@lkb.ens.fr}

\newcommand{\bra}[1]{\ensuremath{\langle #1 |}}
\newcommand{\ket}[1]{\ensuremath{| #1 \rangle}}
\newcommand{\braket}[2]{\ensuremath{\langle #1 | #2 \rangle}}
\newcommand{\matr}[4]{\ensuremath{\left(\begin{array}{cc} #1 & #2 \\ #3 & #4 \end{array} \right)}}
\newcommand{\vectr}[2]{\ensuremath{\left(\begin{array}{c} #1 \\ #2 \end{array} \right)}}
\newcommand{\moy}[1]{\ensuremath{\langle #1 \rangle}}
\newcommand{\op}[1]{\ensuremath{\hat{\mathbf{#1}}}}
\newcommand{\ve}[1]{\ensuremath{#1_1 ... #1_n}}

\begin{abstract}
We present an exhaustive analysis of the light-induced frequency shifts of the photoassociation lines of ultracold metastable $^4He^*$ atoms in a magnetic trap. The measurements of the shifts of several vibrational levels bound in the purely long-range $J=1,~0_u^+$ potential linked to the $2^3S_1-2^3P_0$ asymptote were reported in a previous paper \cite{Kim2005}, and are analyzed here. The simplicity of this system makes it very appropriate for a detailed study. Indeed, the purely long-range character of the excited potential allows one to calculate exact excited molecular wavefunctions and to use asymptotic expansions at large internuclear distances of the ground state wavefunctions appearing in Franck-Condon type integrals. Moreover, the number of collisional channels to be considered is strongly reduced by the absence of hyperfine structure for $^4He^*$ and the use of polarized ultracold atoms and polarized light. This allows us to derive semi-analytical expressions for the shifts showing explicitly their linear dependences on the $s$-wave scattering length $a$ of spin polarized metastable $^4He^*$ atoms. This explains how it is possible to derive the measurement of $a$ from these shifts.
\end{abstract}
\pacs{~32.70.Jz,32.80.Pj,~33.20.Ea,~67.65.+z}
\submitto{\jpb}

\section{Introduction}
Since the breakthrough of the methods for cooling and trapping atoms twenty years ago, a renewal of interest occurred for photoassociation (PA) experiments, in which a pair of colliding cold atoms is promoted into an excited molecular bound state by absorption of laser light. Research in this domain provided a large amount of high precision results in the field of molecular spectroscopy and cold atom collisions \cite{Weiner1999}. When the intensity of the PA laser light is high enough, it is likely to induce frequency shifts which have to be controlled for accurate spectroscopy. Such frequency shifts are interesting processes which result from the perturbation of molecular states by the electromagnetic radiation. They have been studied by several groups dealing with cold alkali atom clouds, both experimentally and theoretically. In particular the cases of Na \cite{McKenzie,Samuelis} and Li \cite{Gerton,Prodan} have been investigated, and the theoretical background has been provided by \cite{BJ,Simoni}. Even if the physics behind these experiments is well understood, a detailed comparison between theory and experiment is somehow limited by the incomplete knowledge and the complexity of the molecular potentials for alkalis. 
 
In the present article we investigate theoretically the light-induced shifts in the photoassociation of ultracold $2^3S_1$ metastable $^4He^*$ atoms. This case is significantly simpler than the case of alkalis for several reasons. First, $^4He^*$ has no hyperfine structure so that the number of channels involved in the calculation is smaller. Second, we consider here photoassociation processes which produce excited molecules in the purely long-range $0_u^+$ potential which can be exactly expressed in terms of atomic parameters (atomic $C_3$ dipole-dipole coefficient and fine structure of the $2^3P$ excited atomic state). The wavefunctions of the excited molecular states which describe giant dimers can thus be exactly calculated \cite{Leonardexp,Venturi2003,Leonardth}. Third, since the wavefunction of the excited molecular state is non-zero only for large values of the internuclear distance $r$, one can use asymptotic expansions of the ground state molecular wavefunctions in the Franck-Condon type matrix elements which appear in the expression of the shifts. We thus do not need to know the wavefunction in the short range domain where it depends critically on the ground state potential. We show in this article that it is then possible to obtain semi-analytical expressions for the light-induced shifts showing explicitly their linear dependence on the $s$-wave scattering length $a$ which describes the collisions between spin polarized ultracold metastable $^4He^*$ atoms.

The results derived in this paper were actually used for the interpretation of a PA experiment with ultracold metastable $^4He^*$ atoms, which was recently published by our group \cite{Kim2005}, and from which we derived a value for $a$. Actually, such light shift measurements were a first step to finding a reliable value for $a$, an important parameter ruling the interatomic interactions and the Bose Einstein condensate properties. The derivation of $a$ from the light shifts was a preliminary study to reaching an even much more accurate value by a second method \cite{Moal2006}. This was  based on two-photon PA spectroscopy and dark resonances, from which we measured the binding energy of the least bound state in the interaction potential of the two metastable atoms. It allowed us to find a value for $a$ at least a hundred times more precise than the best previous determinations, in disagreement with some of them. The preliminary one-photon PA measurement of $a$, derived from the understanding of the light shifts, was very helpful to focus the frequency scan for the final two-photon experiment to a restricted range centred around a reliable value. The interpretation of the physical origin of the light-induced frequency shifts has been briefly explained in the article reporting our experimental results \cite{Kim2005}. The detailed analysis given here is exhaustive and justifies the derivation of the value of $a$ from our previous measurements. 
 
 We first explain in section \ref{sec:origin} the physical origin of the shift of the PA line. We show that it can be related to the well known problem of a bound state coupled to a continuum of states and to a set of bound states \cite{Fano}. In our problem of ultracold atoms held in a magnetic trap, the unique significant entrance collisional channel leading to real photoassociation processes corresponds to two spin polarized atoms colliding in an $s$-wave ($\ell=0$). We point out the two possible origins of the dependence of the shift of the PA line on the scattering length $a$ characterizing the collision in an $s$-wave: dependence on $a$ of the binding energy of the least bound state in the potential of the entrance $s$-wave channel and dependence on $a$ of the Franck-Condon overlaps of the excited and ground molecular wave functions. We emphasize also the importance of taking into account other non $s$-wave ground state channels for evaluating the shift of the PA line. This is due to the fact that the shift is sensitive to non resonant couplings to these channels which may have a non negligible contribution, whereas the photoassociation rate is due to resonant couplings occurring only in the $s$-wave entrance channel. Since the $\ell \neq 0$ ground channels coupled to the excited bound molecular state must satisfy polarization selection rules, one expects a laser polarization dependence of the shifts.
 
 We apply these considerations in section \ref{sec:He} to the photoassociation of two spin polarized $^4He^*$ atoms in a well defined rovibrational state of the excited $J=1,0_u^+$ purely long range potential. Using the selection rules of electric-dipole transitions, we identify the various ground state channels to be considered for two different laser polarizations. We also give the expansion of the excited molecular state in the Hund's case (a) basis which is appropriate to the calculation of the matrix elements of the molecule-laser coupling .
 
 The explicit expression of the shift is derived in section \ref{sec:formulas} from an effective Hamiltonian approach. Two formulas previously obtained by different methods are shown here to be equivalent. In the first one, the shift is given by the well known second-order perturbation expression \cite{Fano,Simoni}. The second one is easier to use since it involves only the wavefunction of the excited state and ground state wave functions at a given scattering energy with well defined boundary conditions \cite{Du,Fedichev1996a,Napolitanoscatt,BJ}. The fact that the excited molecular states in the potential $0_u^+$ are long range allows us to use the asymptotic form of the ground state wavefunction. The second expression of the photoassociative shift derived in section \ref{sec:formulas} can thus be simplified and in section \ref{sec:calc} we derive expressions for the shifts $\delta^v$ of three PA lines ($v=0,1,2$) showing explicit linear dependence of $\delta^v$ on the scattering length $a$. The curves giving $\delta^{v=1}/\delta^{v=0}$ and $\delta^{v=2}/\delta^{v=0}$ as a function of $a$ are given. These theoretical curves were used in \cite{Kim2005} to deduce a range of values of $a$ from our measurements of the shift ratios.

\section{\label{sec:origin}Physical origin of the shift}

\subsection{\label{sec:Fano} Shift of a bound state coupled to a continuum and to a set of bound states}
\begin{figure}
\begin{center}
\includegraphics[scale=0.5]{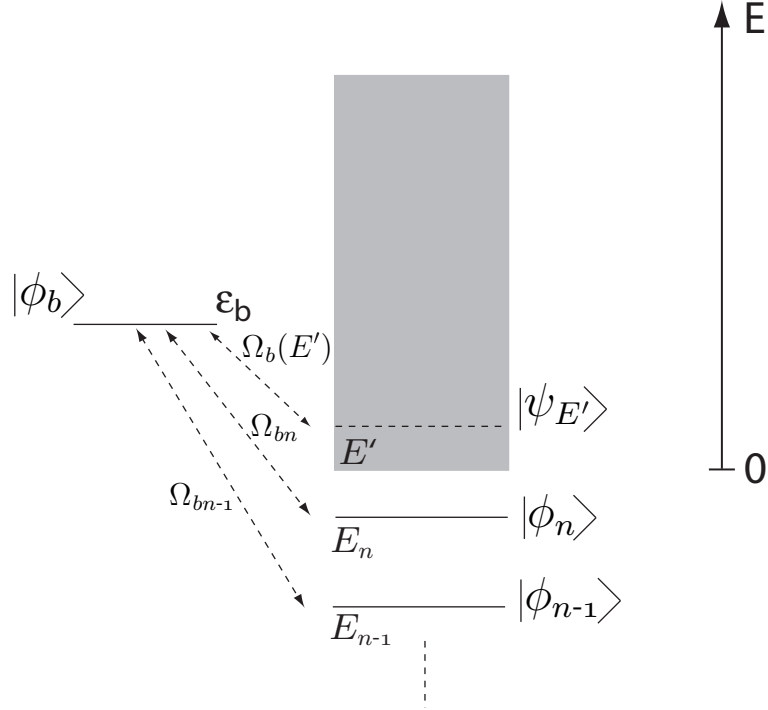}
\caption{\label{fig:Fano}Bound state \ket{\phi_b} of energy $\epsilon_b$ coupled to a set of bound states \ket{\phi_i} of energy $E_i$,~$i=1\dots n$ and to a continuum of states \ket{\psi_{E'}} of energy $0 \leq E' <+\infty$. The matrix element which couples \ket{\phi_b} and a state \ket{\psi_{E'}} of the continuum is noted $\Omega_b(E')$, and the one which couples \ket{\phi_b} and a bound state \ket{\phi_{i}} is noted $\Omega_{bi}$. These couplings induce a shift of the energy $\epsilon_b$ of the excited state $\phi_b$. The continuum of states of energy $E'>\epsilon_b$ gives a contribution to the shift towards negative energies, whereas the continuum of states of energy $E'<\epsilon_b $ and the bound states give a contribution to the shift towards positive energies (see equations (\ref{eq:Fanocont}) and (\ref{eq:Fanobound})).}
\end{center}
\end{figure}

We first recapitulate here on a few results about the coupling of a bound state to a continuum and to a set of bound states \cite{Fano}. Let $\phi_b$ be a bound state of energy $\epsilon_b$, $\phi_i$ a set of $n$ bound states of energy $E_i<0$,~$i=1\dots n$, and $\psi_{E'}$ a continuum of energy normalized states of energy $0<E'<+\infty$ (see Fig \ref{fig:Fano}). We assume that the states are chosen so that the Hamiltonian $H$ is diagonal in the two subspaces spanned by $\phi_b$ on the one hand, and the $\phi_i$ and $\psi_{E'}$ on the other hand :
\begin{eqnarray*}
\bra{\phi_b}H\ket{\psi_{E'}}&=\Omega_b(E') \\
\bra{\phi_b}H\ket{\phi_i}&=\Omega_{bi} \\
\bra{\phi_i}H\ket{\phi_j}&=E_i\delta_{ij} \\
\bra{\psi_{E'}}H\ket{\psi_{E''}}&=E'\delta(E'-E'') \\
\bra{\phi_i}H\ket{\psi_{E'}}&=0
\end{eqnarray*}

We have taken $\hbar=1$. We note for further convenience $\Gamma(E')=|\Omega_b(E')|^2$. The rate $\Gamma(E')$ has the dimension of an energy as the continuum states $\psi_{E'}$ are energy normalized. The probability of excitation of the bound state $\phi_b$ from a continuum state $\psi_E$ exhibits a resonance when $E=\epsilon_b+\delta_b$. The shift $\delta_b=\delta_b^{cont}+\delta_b^{bound}$ is the sum of the contributions $\delta_b^{cont}$ of the continuum states $\Psi_{E'}$ and $\delta_b^{bound}$ of the bound states $\phi_i$.
\begin{equation}
\delta_b^{cont}=\mathcal{PV}\int_0^\infty dE' \frac{\Gamma(E')}{\epsilon_b-E'}
\label{eq:Fanocont}
\end{equation}
\begin{equation}
\delta_b^{bound}=\sum_i \frac{|\Omega_{bi}|^2}{\epsilon_b-E_i}
\label{eq:Fanobound}
\end{equation}
In (\ref{eq:Fanocont}) $\mathcal{PV}$ denotes Cauchy's principal value integral. We find that that the continuum of states of energy $E'>\epsilon_b$ gives a contribution to the shift towards negative energies, whereas the continuum of states of energy $E'<\epsilon_b $ and the bound states $\phi_i$ give a contribution to the shift towards positive energies.

\subsection{Application to photoassociation}
\begin{figure}
\begin{center}
\includegraphics[scale=0.4]{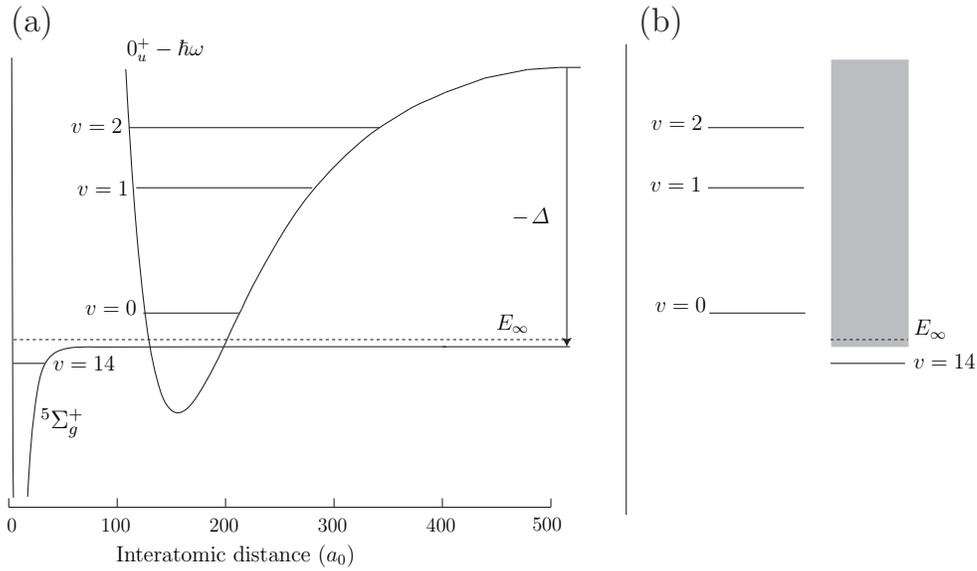}
\caption{\label{fig:PotWave} (a) States and potential curves involved in the experiment of \cite{Kim2005}. Since we are in the dressed-atom picture, the difference between the dissociation energies of the two potential curves is equal to the detuning $\Delta=\omega-\omega_0$, where $\omega/2\pi$ is the laser frequency and $\omega_0/2\pi$ is the frequency of the atomic transition. The spin polarized $^4He^*$ atoms collide with an energy $E_\infty$ in the $^5\Sigma_g^+$ potential, which supports 15 bound states. As the scattering length $a$ for this potential is large and positive, the energy of the least bound state $v=14$ is close to the dissociation threshold. The atoms are photoassociated in bound states $v=0,1,2$ of the purely long-range $J=1,0_u^+$ potential. In this figure, the laser frequency $\omega/2\pi$ is close to resonance with the transition which drives the colliding atoms to the $v=0$ state. (b) Problem in terms of bound states coupled to a continuum and a set of bound states. In the case where the coupling between the ground and excited states is small compared to the level spacing in the excited potential, the influence of $v=1$ and $v=2$ on the photoassociation of $v=0$ can be neglected. This scheme is then analogous to that described in Figure \ref{fig:Fano}. Here $v=0$ is the bound state \ket{\phi_b} which is optically coupled to the scattering states \ket{\psi_E'} and to the bound state  \ket{\phi_{i=14}}  of the $^5\Sigma_g^+$ potential.}
\end{center}
\end{figure}

We now apply the previous considerations to the photoassociation (PA) of a pair of colliding atoms in a molecular bound state of energy $E_b$ in an excited potential. In the light-dressed picture, the excited potential is shifted by an amount $-\omega$ (we remind that $\hbar=1$) corresponding to the energy of a photon of the PA laser. The problem is analogous to the one of a bound state coupled to a continuum and to a set of bound states (see Fig. \ref{fig:PotWave} in  the metastable helium case). Using the notations of paragraph \ref{sec:Fano}, $\phi_b$ is the excited bound state of energy $\epsilon_b=E_b-\omega$, the states $\psi_{E'}$ and $\psi_i$ correspond to the scattering and bound states in the ground state potential, $\Omega_b(E')$ and $\Omega_{bi}$ are the light-induced couplings between the ground and excited states, which are proportional to the square-root of the PA laser intensity. We assume that these couplings are weak enough so that $\phi_b$ can be considered isolated from the other states in the excited potential. We also assume that only the least bound state of energy $E_{n}$ in the ground state potential, which is the closest to resonance, has a non negligible influence on the spectroscopy of the excited state. In the context of ultracold collisions between atoms, the energy of the initially populated scattering state $E_\infty$ is close to the dissociation  threshold of their interaction potential, which means $E_\infty\simeq 0$. The unperturbed resonance condition $E_\infty=\epsilon_b$ imply that $\epsilon_b\simeq 0$. Using these simplifications, the contributions in equations (\ref{eq:Fanocont}) and (\ref{eq:Fanobound}) give for the total shift:

\begin{equation}
\delta_b\sim-\mathcal{PV}\int_0^{+\infty} dE' \frac{\Gamma(E')}{E'}-\frac{\Omega_{bn}^2}{E_{n}}
\label{eq:introshift}
\end{equation}

As $\Gamma(E')$ and $\Omega_{bn}^2$ are proportional to the laser intensity, the shift $\delta_b$ is also proportional to the laser intensity.

\subsection{Coupling with a ground state $s$-wave channel}
Considering the coupling with an $s$-wave collisional channel characterized by a scattering length $a$, two origins can be identified in equation (\ref{eq:introshift}) for the dependence of the shift on $a$. 

First, in case of a large and positive scattering length $a$, the binding energy $E_n$ of the least bound state is approximated by $-1/2\mu a^2$, where $\mu$ is the reduced mass of the colliding system. As noticed in \cite{Simoni}, and also mentioned above, the contributions from the bound state and from the continuum of states have opposite signs in equation (\ref{eq:introshift}): at zero scattering energy, the continuum contribution is negative, whereas the bound state one is positive, and all the more important that the scattering length $a$ is large. The shift is sensitive to the vicinity of the least-bound state in the ground state interaction potential and thus to $a$. 

\begin{figure}
\begin{center}
\includegraphics[scale=0.8]{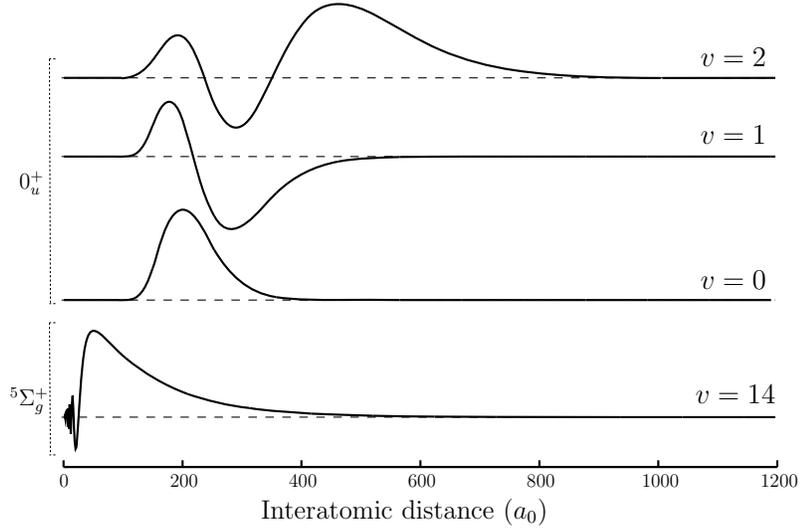}
\caption{\label{fig:Potpot}Wavefunctions $v=14$ for the molecule in the ground state potential $^5\Sigma_g^+$, and $v=0,1,2$ for the molecule in the excited purely long-range $J=1,0_u^+$ potential. The overlap of the wavefunctions $v=14$ with $v=0,1,2$ appears in $\Omega_{b(i=14)}$ in the bound states contribution (\ref{eq:Fanobound}) to the shifts of $v=0,1,2$. The wavefunction of $v=1$ crosses zero an odd number of times whereas the wavefunctions for $v=0$ and $v=2$ cross zero an even number of times. The wavefunction of $v=14$ varies monotonously in the range of interatomic distances where the wavefunctions of $v=0,1,2$ take non negligible values. As a consequence, in case of the overlap with $v=1$, positive and negative contributions approximately cancel each other, and the contribution of $v=14$ to the shift of $v=1$ is expected to be smaller than its contribution to the shift of $v=0$ and $v=2$.}
\end{center}
\end{figure}

The second origin for the dependence of the shift on $a$ is found in the stimulated rates $\Omega_{bn}^2$ and $\Gamma(E')$ which contain overlap integrals involving $a$-dependent wavefunctions for the electronic ground-state and wavefunctions for the excited molecular state. The shift for each excited molecular state depends differently on $a$. Figure \ref{fig:Potpot} shows for example the wavefunctions of the ground and excited molecular bound states involved in the experiment described in \cite{Kim2005}. Due to the fact that the molecules in the excited electronic state lie in a purely long range potential, the nodes of their  wavefunctions are localized at a distance corresponding to the vanishing exponential tail of the molecular wavefunction in the electronic ground state. As mentioned in \cite{Simoni}, if the wavefunction of the excited bound state changes sign an odd number of times, positive and negative contributions  approximately cancel each other in the bound-bound overlap integral contained in $\Omega_{bn}$ , and the contribution (\ref{eq:Fanobound}) to the shift is smaller than in the case of an even number of changes of sign of the wavefunction. Therefore, the investigation of the shift for several excited states provides complementary information on the scattering length.

\subsection{Contribution of other ground state channels}
At very low scattering energy $E_\infty$, atoms colliding in a $\ell\neq 0$ partial wave are held apart by the centrifugal barrier $\ell(\ell+1)/2\mu r^2$, which prevents them from being efficiently photoassociated. The free-bound stimulated rate $\Gamma_\ell(E_\infty)>0$  for an incoming $\ell>0$ partial wave can be neglected compared to $\Gamma_{\ell=0}(E_\infty)$. Only $s$-wave collisional channels contribute to resonant photassociation processes producing excited molecules. However, the excited level shift ($\ref{eq:introshift}$) involves the contribution of the whole spectrum. States of higher energy $E'$, for which $\Gamma_{\ell\neq 0}(E')$ is not negligible compared to $\Gamma_{\ell=0}(E')$, can contribute noticeably to the shift even though they are non resonant. For example, in a previous experiment with a Bose-Einstein condensate of Na \cite{McKenzie}, the $d$-wave contribution is enhanced by a shape resonance and is much larger than the $s$-wave contribution.

Owing to the contribution of non $s$-wave collisions, the shift is expected to depend on the polarization of light, as already mentioned in \cite{Montalvao}. The collisional channels which are coupled to the excited state depend on the polarization of light. Therefore the investigation of the shift as a function of the light polarization allows one to tune the relative contribution of the $s$-wave collisional channel, which depends on $a$, and of the other $\ell\neq 0$ contributions. 

As a consequence, when investigating the light shift of the excited state, one has to carefully take into account all the ground state collisional channels which are coupled, and not only those which are resonant.

\section{\label{sec:He}Helium interaction potentials}{\label{sec:Hepot}}

\subsection{\label{sec:collme}Collisions between metastable helium atoms}
\begin{figure}
\begin{center}
\includegraphics[scale=0.3]{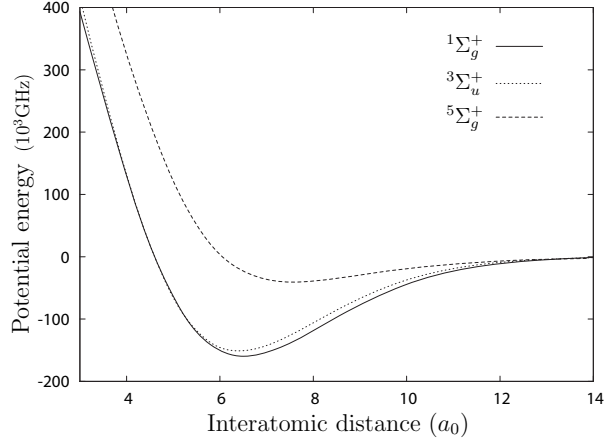}
\caption{\label{fig:Muller} Potential energy curves involved in the collision between metastable helium atoms as calculated by \cite{Muller,Starckmeyer,Gadea2002,Gadea2004,Przybytek}.}
\end{center}
\end{figure}

In the absence of a light field and neglecting the weak spin-spin interaction between atoms, the total electronic spin $S$ and the relative angular momentum $\ell$ of two spin-1 colliding $^4He^*$ atoms are conserved to good approximation. A basis to describe the collision of a pair of colliding atoms $a$ and $b$ is given by (\ref{eq:SMSlMlbasis}), where $S_a$, $S_b$ are the spins of atom $a$, atom $b$, $\ell$ the relative rotational angular momentum, $M_S$ and $M_\ell$ are the projections of $S$ and $\ell$ on the lab-fixed quantization axis. They have no electronic orbital angular momentum as the atoms are in a $2^3S$ state. 
\begin{equation}
\ket{(S_a=1,S_b=1)S,M_S;\ell,M_\ell}
\label{eq:SMSlMlbasis}
\end{equation}
As shown in \ref{sec:appsym}, due to the bosonic nature of the metastable helium atoms, singlet ($S=0$) and quintet ($S=2$) spin states collide in even $\ell$ partial waves, whereas triplet ($S=1$) spin states collide in odd $\ell$ partial waves. Due to the bosonic nature of both the helium atom and helium nucleus, singlet and quintet (triplet) states have the \emph{gerade} (resp. \emph{ungerade}) symmetry under the inversion of the electronic coordinates through the center of charge of the molecule. As both atomic $2^3S_1$ orbitals are spherical, the spatial part of the electronic wavefunction is not changed by a symmetry through a plane containing the intermolecular axis, and the colliding pair has a $+$ symmetry. The colliding states are $\Sigma$ states because the colliding metastable helium atoms have no electronic orbital angular momentum. Figure \ref{fig:Muller} shows the interaction potentials $^1\Sigma_g^+$,$^3\Sigma_u^+$ and $^5\Sigma_g^+$  associated to the quintet, triplet and singlet states respectively.

\subsection{\label{sec:ground} Ground states optically coupled}

In the experiment described in \cite{Kim2005}, the $^4He^*$ atoms are spin-polarized and held in a magnetic trap. The atoms collide at energies such that only the $s$-wave part of the scattering wave contributes to the photoassociation, and the state which characterizes the internal and rotational parts of the system is $\ket{(S_a=1,S_b=1),S=2,M_S=2;\ell=0,M_\ell=0}$. The total angular momentum $J$ is defined as $S+L+\ell$, where $L$ is the total electronic orbital angular momentum and $L=0$ for a pair of metastable atoms. Therefore the colliding state is a $J=2,M_J=2$ state.

The total angular momentum $J$ is not conserved in presence of a laser field. We note $J'$ and $M_{J'}$, the total angular momentum of the excited state and its projection on the lab-fixed quantization axis. We remind here the selection rules  for electric-dipole transitions :
\begin{itemize}
\item $|J-1|\leq J'\leq|J+1|$
\item $M_{J'}-M_J=q$ where $q=-1,0,1$ for $\sigma^-$,$\pi$ and $\sigma^+$ polarization respectively
\item there is no coupling between states of the same symmetry under the inversion of the electronic coordinates through the center of charge of the quasi molecule
\end{itemize}

In the experiment described in \cite{Kim2005}, the atoms are excited in vibrational levels of the purely long-range $J'=1,M_{J'}=1;0_u^+$ potential. As already mentioned at the end of section \ref{sec:origin}, the initial scattering state $\ket{(S_a=1,S_b=1),S=2,M_S=2;\ell=0,M_\ell=0}$ is not the only ground state which will be relevant for evaluating the light shift. Once the excited $0_u^+$ state is formed, it can be optically coupled to other ground states which contribute to the light shift. Because of the selection rules, the projection of the total angular momentum in the ground state is $M_J=M_S+M_\ell$ and has to be $M_J=1-q$. Furthermore, $S$ and $\ell$ should be such that they can form a total angular momentum $J$ that satisfies the triangular inequalities with $J'=1$ and the photon angular momentum equal to 1. Therefore, the properly symmetrized states (see (\ref{eq:Slbasis}) in \ref{sec:appsym}) which are coupled to the excited state by $\sigma^-$ and $\sigma^+$ light are given by (\ref{eq:coupledminusstates}) and (\ref{eq:coupledplusstates}) respectively using the simplified notation \ket{S,M_S;\ell,M_\ell} (see (\ref{eq:Slbasis2})):
\begin{equation}\left\{\begin{array}{ccccc} 
\ket{22;00}&~&~&~&~ \nonumber \\ 
\ket{22;20}&\ket{21;21}&\ket{20;22}&\ket{00;22}&~ \nonumber \\ \ket{22;40}&\ket{21;41}&\ket{20;42}&\ket{2-1;43}&\ket{2-2;44} \label{eq:coupledminusstates} 
\end{array}\right.\end{equation} 
\begin{equation}\left\{\begin{array}{cccccc} 
\ket{20;00}&\ket{00;00}&~&~&~&~ \nonumber \\
\ket{20;20}&\ket{21;2-1}&\ket{22;2-2}&\ket{2-1;21}&\ket{2-2;22}&\ket{00;20} \nonumber \\
\ket{20;40}&\ket{21;4-1}&\ket{22;4-2}&\ket{2-1;41}&\ket{2-2;42}&~ \label{eq:coupledplusstates}
\end{array}\right.\end{equation}
Several remarks can be formulated :
\begin{itemize}
\item In (\ref{eq:coupledminusstates}) $M_S+M_\ell=2$ and in (\ref{eq:coupledplusstates}) $M_S+M_\ell=0$. 
\item The scattering wavefunctions for the channels $\ket{22;00}$ and $\ket{20;00}$ which appear in both (\ref{eq:coupledminusstates}) and (\ref{eq:coupledplusstates}) depend on the quintet scattering length $a$ which is searched in \cite{Kim2005}. The wavefunctions in the other channels do not depend on $a$, because they correspond to either a singlet state, or to $d$- or $g$-wave collisions.
\item By changing the polarization of light, one can change the relative contribution of $a$-dependent and $a$-independent channels as the coupled channels are different in both cases.
\item The state $\ket{00;00}$ involves the scattering length of the $^1\Sigma_g^+$ potential (see Fig. \ref{fig:Muller}), which is estimated to be $a_1=35\pm15~a_0$ according to previous calculations \cite{Muller,Leocollision}.
\end{itemize} 

\subsection{\label{sec:excited} Description of the excited state}
Contrary to the case of molecular potentials connected to the $S+S$ asymptote investigated in paragraph \ref{sec:ground}, $S$ and $\ell$ are no longer good quantum numbers to describe the molecular potentials which dissociate into the $S+P$ asymptote. Due to the fine structure splitting of the $P$ state, the spin $S$ is mixed to the orbital angular momentum $L$ and is not conserved. The interaction between the $S$ and $P$ atoms is not isotropic as it depends on the relative orientation of the electronic orbital angular momentum and the molecular axis, and $\ell$ is not conserved. The remaining good quantum numbers to characterize the excited state are $J$,$M_J$ and $\Omega$, the projection of $L+S$ on the molecular axis. The purely long-range $0_u^+$ potential used in \cite{Kim2005} was previously investigated in great detail experimentally and theoretically \cite{Leonardexp,Venturi2003,Leonardth}. We recapitulate here the essential points needed for our purpose. 

The purely long-range $0_u^+$ potential which connects to the $2^3S_1+2^3P_0$ asymptote is mainly determined by the atomic $C_3$ dipole-dipole coefficient and the fine-structure splitting of the $2^3P$ state. It is shallow (about 2 GHz deep) and supports five bound states which have inner turning points at internuclear separations of approximately $150 \mathrm{a_0}$. Their  electronic state is close to that of two separated $2^3S$ and $2^3P_0$ atoms.
The decomposition of the electronic $0_u^+$ state in the Hund's case (a) basis is needed to estimate the light-induced coupling between the ground and excited states.  It has contributions from the states $^5\Pi_u$,$^3\Pi_u$,$^5\Sigma_u^+$ and $^1\Sigma_u^+$. The $^3\Sigma_u^+$ state is not present as it belongs to the $0_u^-$ subspace. We write the decomposition of the purely long-range $\ket{0_u^+}$ electronic state in the Hund's case (a) basis in equation (\ref{eq:zerou}), where the $r$-dependent $\alpha$ coefficients are obtained by the diagonalisation of the $r$-dependent electronic Hamiltonian \cite{Leonardth}:

\begin{equation}
\ket{0_u^+}=\alpha_{^5\Sigma_u^+}(r)\ket{^5\Sigma_u^+}+\alpha_{^5\Pi_u}(r)\ket{^5\Pi_u}+\alpha_{^3\Pi_u}(r)\ket{^3\Pi_u}+\alpha_{^1\Sigma_u^+}(r)\ket{^1\Sigma_u^+}
\label{eq:zerou}
\end{equation}

Experimentally, the rovibrational $J=1$ states are probed, where $J$ is the total angular momentum of the molecule. The $J=2$ states do not exist as they are not symmetric under the exchange of the two bosonic helium nuclei, and the $J=3$ states could not be observed experimentally because of the very weak coupling with the initial state. This can be understood by the fact that the transition is forbidden in the limit of two non-interacting atoms and that the electronic state of the excited long range molecule is close to the one of two non interacting atoms.  As the $J=1$ states are experimentally  produced from a $J=2$ state (see subsection \ref{sec:ground}), owing to selection rules, the projection $M_J$ of $J$ on the molecular axis is $M_J=1$. Due to rotational couplings, more Hund's case (a) states than the four previous one are involved in the decomposition of the photoassociated levels, but their contribution is only non negligible for interatomic distances where the purely long-range $0_u^+$ potential crosses other adiabatic potentials. This occurs at interatomic distances well below the inner turning point of the photoassociated levels. Therefore, we keep only the four electronic states (\ref{eq:zerou}) in the expression of the wavefunction for the rotating molecule.
\begin{equation}
\ket{J'=1, M_{J'}=1;0_u^+}= \Sigma_i \alpha_i(r)\ket{J'=1,M_{J'}=1,i}
\label{eq:zerouj}
\end{equation}  
where $i={^5\Sigma_u^+,^5\Pi_u,^3\Pi_u,^1\Sigma_u^+}$. In \ref{sec:couplings}, we give the explicit expression (\ref{eq:touz2casa}) of the states \ket{J'=1,M_{J'}=1,i} in terms of states of the Hund's case (a) basis for a rotating molecule $\ket{J, M_J,\Omega,S,\Sigma,\Lambda,w}$.

\section{\label{sec:formulas}Two equivalent formulas for the molecular light-shift}
We derive here formulas for the shifts of the bound levels in the electronic excited state \ket{J'=1,M_{J'}=1;0_u^+} described in section  \ref{sec:excited}, due to the optical coupling with ground state channels described in section \ref{sec:ground}. Using an effective Hamiltonian treatment and two different expressions for the field-free propagator for the ground state channels, we show the equivalence between two formulas established in previous papers. The first one involves a sum over the whole energy spectrum in the electronic ground state and was derived in \cite{Fano} and in \cite{Simoni} in the context of photoassociation. The second one derived by different means in \cite{Du,Fedichev1996a,Napolitanoscatt,BJ} involves only ground state wavefunctions at a given scattering energy, and is therefore much easier to estimate numerically. The equivalence between the two formulas which we demonstrate here can be regarded as an extension of the sum rule described in \cite{Dalgarno}. Let us emphasize that the contribution of the bound states to the shift, which is explicit in the first formula, is also included in the second one.

We now present the effective Hamiltonian treatment used to derive the shifts. Given that the couplings between the ground state channels due to the very weak spin-dipole interaction \cite{Shlyapnikov1994} can be neglected, we restrict ourselves to a two-channel treatment between a $\ket{g}=\ket{S,M_S;\ell,M_\ell}$ ground state and the $\ket{e}=\ket{J'=1,M_{J'}=1;0_u^+}$ excited state. This approximation will be discussed at the end of this section. The shifts $\delta_{S,M_S;\ell,M_\ell}$ calculated in this way for the states (\ref{eq:coupledminusstates}),(\ref{eq:coupledplusstates}) can then be added up to calculate the total shift $\delta$:
\begin{equation}
\delta=\sum_{g=S,M_S;\ell,M_\ell}\delta_{g}
\label{eq:sumshift}
\end{equation}
The coupling induced by the laser of intensity $I$ between a ground channel $g=(S,M_S;\ell,M_\ell)$ and the excited channel $e$ is defined as  $\Omega_{S,M_S;\ell,M_\ell}(r)=\Omega_{eg}(r)$ and is calculated in appendix B. Note that $\Omega_{eg}(r)$ is obtained after an integration over electronic variables and over the angular variables of the molecular axis so that $\Omega_{eg}(r)$ acts only on the variable $r$, distance between the two nuclei. The problem can be described by a $r$ dependent two-channel Hamiltonian $H_{2C}$ (\ref{eq:H2C}) :
\begin{eqnarray}
H_{2C}&=&H_0+V_R \label{eq:H2C} \\ H_0&=&-\mathbf{1}\frac{1}{2\mu}\frac{d^2}{dr^2}+\matr{V_{g}(r)}{0}{0}{V_{e}(r)-\Delta-i\frac{\gamma}{2}} \nonumber \\
V_R&=&\matr{0}{\Omega_{ge}(r)}{\Omega_{eg}(r)}{0} \nonumber
\end{eqnarray}
In (\ref{eq:H2C}) $V_g(r)$ is the ground state potential which is the sum of the interaction potential  $^{2S+1}\Sigma_g^+$ and of the centrifugal potential $\ell(\ell+1)/2\mu r^2$. $V_e(r)$ is the $J=1,0_u^+$ potential. $\mathbf{1}$ is the identity matrix. $\gamma$ accounts for spontaneous emission from the excited state. We choose the reference of energy at the dissociation energy of the potential associated to the channel $g:(S=2,M_S=2;\ell=0,M_\ell=0)$, which corresponds to the collision between spin polarized atoms. $\Delta$ is the detuning $\omega-\omega_0$ between the laser and the atomic frequencies (see Figure \ref{fig:PotWave}(a)). In presence of a magnetic field $B$, the potentials corresponding to channels with different projections $M_S$ are shifted proportionally to $g_S\mu_B B$, where $g_S$ is the electron spin gyromagnetic ratio and $\mu_B$ the Bohr magneton. We define therefore the energy $E_{M_S}^\infty$ and the wavevector $k_{M_S}^\infty$ which will appear in the matrix elements of the effective Hamiltonian $H_{eff}(E)$ :

\begin{eqnarray}
E_{M_S}^\infty=V_g(\infty)=(M_S-2)g_S\mu_BB \label{eq:Einf} \\
k_{M_S}^\infty=\sqrt{2\mu(E-E_{M_S}^\infty)} \label{eq:kinf}
\end{eqnarray}

To obtain the effective Hamiltonian, we define $P$ and $Q$ as the projectors onto the states in the ground channel and the excited channel respectively (\ref{eq:PQ}). These states can be characterized by the quantum numbers $\alpha$ and $\beta$ which denote either a vibrational number or the scattering energy of the eigenstates of $PH_0P$ and $QH_0Q$.
\begin{eqnarray}
P&=&\sum_{\alpha}\ket{g,\alpha}\bra{g,\alpha} \nonumber \\
Q&=&\sum_{\beta}\ket{e,\beta}\bra{e,\beta} \label{eq:PQ}
\end{eqnarray}
Defining the field free propagator at energy $E$ with outgoing boundary conditions $G_0^{+}(E)=(E-H_0+i\epsilon)^{-1}$ (chapter XIX in \cite{Messiah}), we obtain the effective Hamiltonian $H_{eff}$ for the excited state:
\begin{equation}
H_{eff}(E)=-\frac{1}{2\mu}\frac{d^2}{dr^2}+V_{e}(r)-(\Delta+i\frac{\gamma}{2})+QV_RPG_0^{(+)}(E)PV_RQ
\label{eq:Heff}
\end{equation} 

From now on, we will consider that only one bound state of vibrational number $v$ of the excited potential is efficiently coupled by light, which corresponds to the isolated level approximation. Its validity will be discussed later. We have therefore $Q=Q_v=\ket{e,v}\bra{e,v}$. The matrix element $\bra{e,v}H_{eff}\ket{e,v}$ gives the energy of the bound state $v$ perturbed by light. The following expression which appears in $\bra{e,v}H_{eff}\ket{e,v}$ contains a real part $\delta^v(E)$ which corresponds to the light-induced energy shift of the level $v$, and an imaginary part $\Gamma^v(E)$ corresponding to the light-induced broadening:
\begin{equation}
\bra{e,v}Q_vV_RPG_0^{(+)}(E)PV_RQ_v\ket{e,v}=\delta^v_g(E)-i\Gamma^v_g(E)/2
\label{eq:shiftbroad}
\end{equation} 
Let us consider the matrix element $\bra{r'}PG_0^{(+)}(E)P\ket{r}=\mathcal{G}^+(E;r,r')$ of the propagator $PG_0^{(+)}(E)P$. It is related to the matrix element of (\ref{eq:shiftbroad}) by equation (\ref{eq:shiftbroad2}), noting $\phi^e_v(r)$ for the wavefunction associated to the state \ket{e,v} which is real. We use also the fact that $\Omega_{eg}$ is real so that $\Omega_{eg}=\Omega_{ge}$.
\begin{equation}
\fl\bra{\phi^e_v}QV_RPG_0^{(+)}(E)PV_RQ\ket{\phi^e_v}=\int_0^\infty\int_0^\infty \Omega_{eg}(r)\Omega_{eg}(r')\mathcal{G}^+(E;r,r')\phi^v_e(r)\phi^v_e(r')dr dr'
\label{eq:shiftbroad2}
\end{equation}
$\mathcal{G}^+(E;r,r')$ can be written in two following ways :
\begin{enumerate}
\item The definition $G_0^{(+)}(E)=(E+i\epsilon-H_0)^{-1}$ of the propagator and the expression of $PG_0^{(+)}(E)P$ in terms of discrete eigenstates $\ket{g,i=1\dots n}$ of energy $E^n_g$, and scattering eigenstates $\ket{g,E'}$ of energy $E'$ in the channel $g$ provide (\ref{eq:SimoniGreen}). In (\ref{eq:SimoniGreen}), $\mathcal{PV}$ denotes Cauchy's principal value, $\phi^i_g(r)$ is the real wavefunction associated to the bound level $\ket{g,i=1\dots n}$, $\phi^E_g(r)$ is the energy normalized real wavefunction associated to the scattering state $\ket{g,\alpha=E}$ and $E_g^\infty$ the dissociation energy of the potential associated with the channel $g$ (see (\ref{eq:Einf})).
\begin{equation}
\fl\mathcal{G}^+(E;r,r')=\sum_{i=1}^n \frac{\phi^i_g(r')\phi^i_g(r)}{E-E^i_g}+\mathcal{PV}\int_{E_g^\infty}^{+\infty}  \frac{\phi^{E'}_g(r')\phi^{E'}_g(r)}{{E-E'}}dE'-i\pi\phi^E_g(r')\phi^E_g(r)
\label{eq:SimoniGreen}
\end{equation} 
The wavefunction $\phi^E_g$ is the solution of the radial Schr\"odinger equation associated with the channel $g$, and has the following boundary condition for $r\rightarrow\infty$, which involves the phase shift $\eta_{g}(k^\infty_{g})$ and the wavevector $k^\infty_{g}$
\begin{equation}
\phi^{E}_g(r\rightarrow\infty)=\sqrt{\frac{2\mu}{\pi k^\infty_{g}}}\sin \left[ k^\infty_{g} r-\ell\frac{\pi}{2}+\eta_{g}(k_g^\infty) \right] 
\label{eq:regular}
\end{equation}

\item The solution of the differential equation which defines the propagator $\mathcal{G}^+(E;r,r')$  with proper boundary conditions gives (chapter XIX in \cite{Messiah}) : 
\begin{equation}
\mathcal{G}^+(E;r,r')=-\pi\phi^E_g(r_<)\phi^{E+}_{g}(r_>)
\label{eq:MessiahGreen}
\end{equation} 
$r_<$ and $r_>$ are the smaller and the larger of $r$ and $r'$. $\phi^{E+}_{g}$ is the solution of the same Schr\"odinger equation as $\phi^{E}_{g}$, but with outgoing boundary conditions at infinity:
\begin{equation}
\phi^{E+}_{g}(r\rightarrow\infty)= \sqrt{\frac{2\mu}{\pi k^\infty_{g}}}\exp \left[i\left( k^\infty_{g} r-\ell\frac{\pi}{2}+\eta_{g}(k^\infty_{g}) \right)\right] \label{eq:outgoing}
\end{equation}
\end{enumerate} 

Inserting (\ref{eq:SimoniGreen}) and (\ref{eq:MessiahGreen}) into (\ref{eq:shiftbroad2}), and identifying the real and imaginary part of (\ref{eq:shiftbroad2}) with the ones of (\ref{eq:shiftbroad}), we get the light-induced broadening and shift. Both (\ref{eq:SimoniGreen}) and (\ref{eq:MessiahGreen}) give the same form for the light-induced broadening $\Gamma^v(E)$:
\begin{equation}
\Gamma^v_g(E)=2\pi\left|\int_0^\infty{\phi^E_g(r)\Omega_{eg}(r)\phi^v_e(r)dr}\right|^2
\label{eq:Gamma}
\end{equation} 

The expression for the shift $\delta^v(E)$ is different whether equation (\ref{eq:SimoniGreen}) or (\ref{eq:MessiahGreen}) is used, giving respectively (\ref{eq:SimoniShift}) and (\ref{eq:MessiahShift}):

\begin{equation}
\fl\delta^v_g(E)=\sum_{i=1}^n \frac{\left|\int_0^\infty{\phi^i_g(r)\Omega_{eg}(r)\phi^v_e(r)dr}\right|^2}{E-E^i_g}+\mathcal{PV} \int \frac{\left|\int_0^\infty{\phi^{E'}_g(r)\Omega_{eg}(r)\phi^v_e(r)dr}\right|^2}{E-E'}dE' \label{eq:SimoniShift} 
\end{equation}
\begin{equation}
\delta^v_g(E)=2\pi\int_{0}^\infty{dr \int_{0}^r{dr'\Omega_{eg}(r) \Omega_{eg}(r')\tilde{\phi}^E_g(r) \phi^E_g(r') \phi^v_e(r) \phi^v_e(r')}} \label{eq:MessiahShift}
\end{equation}

Here, $\tilde{\phi}^{E}_g$ is the energy normalized solution of the radial Schr\"odinger equation at energy $E$ for the channel $g$, which is related to the real part of (\ref{eq:outgoing}) :
$$\tilde{\phi}_{E}^g(r\rightarrow\infty)= -\sqrt{\frac{2\mu}{\pi k^\infty_{g}}}\cos \left( k^\infty_{g} r-\ell\frac{\pi}{2}+\eta_{g}(k_g^\infty) \right)$$

The expression (\ref{eq:MessiahShift}), which was derived by different ways in \cite{Du,Fedichev1996a,Napolitanoscatt,BJ} is easier to use than the equation (\ref{eq:SimoniShift}) as it involves only wavefunctions at a given scattering energy $E$ rather than at all energies $E'$ and $E^{i}_g$. Both formulas are derived from two exact expressions of the field-free propagator (\ref{eq:SimoniGreen}) and (\ref{eq:MessiahGreen}), and are thus equivalent \cite{Dalgarno}. In particular, (\ref{eq:MessiahShift}) includes the contributions of the bound states to the shift, which are explicit in (\ref{eq:SimoniShift}).

Two approximations were done here. First, the isolated level approximation : when the PA laser is close to resonance
with an excited bound level, the influence of the other excited levels is neglected. It is valid as far as the laser intensity is such that $\delta^v(E)$ and $\Gamma^v(E)$ are small compared to the excited level spacing  \cite{Du,Fedichev1996a,Napolitanoscatt}. If this approximation breaks down, the level shift is no more a quadratic function of the couplings $\Omega_{eg}(r)$, and thus no more proportional to the laser intensity \cite{Du}. We are only  interested here in shifts linear with laser intensity as observed in the experiment described in \cite{Kim2005}. Second, we neglected the spin-dipole interaction between ground state channels. Taking them into account would involve operators $Q_vV_RP'G_0^{(+)}(E)PV_RQ_v$ in the effective Hamiltonian (\ref{eq:Heff}), where $P'$ is defined similarly as in  (\ref{eq:PQ}) but with a different ground state channel $g'$. The matrix elements of these operators would have to be included in the shift calculation. Their importance compared to the non neglected matrix elements of $Q_vV_RPG_0^{(+)}(E)PV_RQ_v$ is related to the probability of spin relaxation compared to the probability of elastic collision per collision event. These probabilities were calculated in \cite{Shlyapnikov1994,Fedichev1996b,Venturiclosecoupled} and indicate that the matrix elements we neglected are four orders of magnitude smaller than those involved in our calculation of the shift.

As a final comment, we notice that the Hamiltonian (\ref{eq:H2C}) describes an optical Feschbach resonance \cite{Fedichev1996a}. The two potentials are shifted from each other using a laser field, similarly to what is achieved with a magnetic field to tune the scattering lengths \cite{Inouye}. The use of light coupling to tune the scattering length has already been investigated both theoretically \cite{Fedichev1996a,Napolitanoscatt,BJscatt} and experimentally \cite{Fatemi,Theis}. The case of metastable helium was considered in \cite{Koelemeij}. Here, the shift is expressed as a function of the unperturbed wavefunctions, and will be evaluated as a function of the unperturbed scattering length in section \ref{sec:calc}.

\section{\label{sec:calc}Calculations of the shift for Helium}

\begin{figure}
\begin{center}
\includegraphics[scale=0.5]{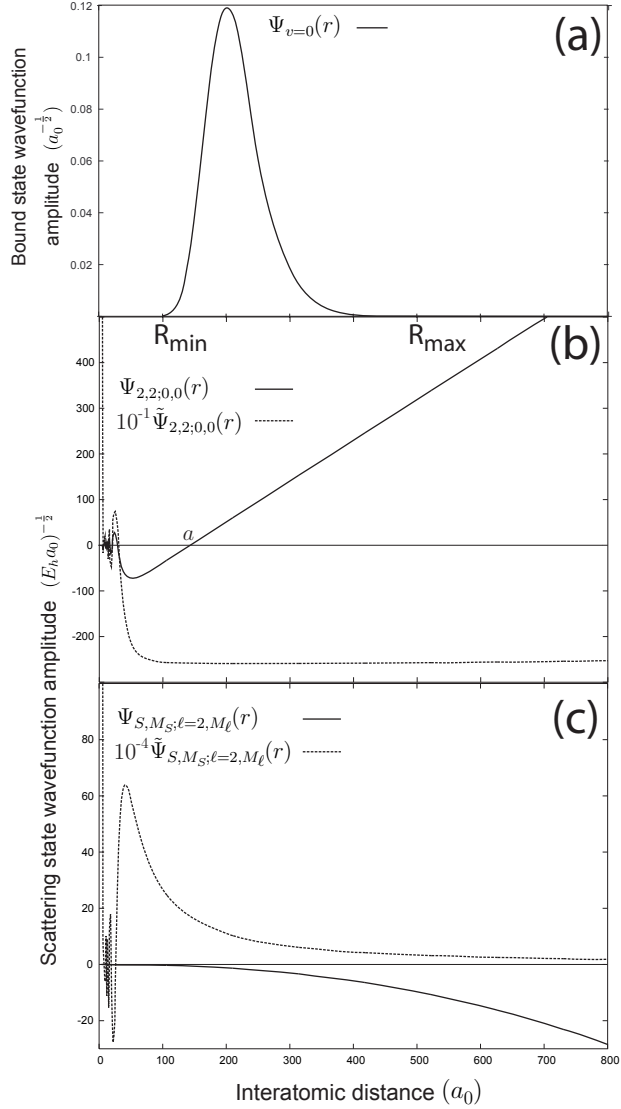}
\caption{\label{fig:wf} Wavefunctions involved in the calculation of the shift (\ref{eq:MessiahShift}). Inset (a) shows the wavefunction of the excited level $v=0$ in the $J=1,0_u^+$ potential which takes non negligible values between $R_{min}=100~a_0$ and $R_{max}=600 ~a_0$. Inset (b) shows the regular and irregular wavefunctions $\psi_{2,2;0,0}$ and $\tilde{\psi}_{2,2;0,0}$ for the collision in an $s$-wave scattering channel at low scattering energy $E_\infty=(k^{\infty}_{M_S=2})^2/2\mu$. The wavevector $k^{\infty}_{M_S=2}$ is such that $k^{\infty}_{M_S=2} r <<1$ for $R_{min}\leq r \leq R_{max} $. As a consequence, for this range of values of $r$ $\psi_{2,2;0,0}(r)\propto\sin\left(k^\infty_{M_S=2}(r-a)\right)\simeq k^\infty_{M_S=2}(r-a)$ and $\psi_{2,2;0,0}(r)\propto-\cos\left(k^\infty_{M_S=2}(r-a)\right)\simeq-1$. Inset (c) shows the regular and irregular wavefunctions $\psi_{S,M_S;\ell=2;M_\ell}$ and $\tilde{\psi}_{S,S_S;\ell=2,M_\ell}$ for the collision in a $d$-wave scattering channel. In the range $R_{min}\leq r \leq R_{max} $, the interaction between the atoms colliding in a $\ell$ wave is mainly due to the centrifugal potential $\ell(\ell+1)/2\mu r^2$. $\psi_{S,M_S;\ell=2,M_\ell}$ and $\tilde{\psi}_{S,M_S;\ell=2,M_\ell}$ are well approximated using spherical Bessel functions of the first and second kind (see equations (\ref{eq:dgf}) and (\ref{eq:dgg})). It can be noticed that the product of the regular and irregular wavefunctions which appears in (\ref{eq:MessiahShift}) is of the same order of magnitude in (b) and (c). This shows that the $d$-wave contribution to the shift may not be negligible compared to the $s$-wave one.}
\end{center}
\end{figure}

The expression of the light-induced energy shift in equations (\ref{eq:sumshift}) and (\ref{eq:MessiahShift}) involves the overlap of the regular and irregular scattering wavefunctions with the wavefunction of the excited bound levels. As these excited bound levels lie in a purely long-range potential, the integrals do not depend on the short-range part of the ground state wavefunctions, and numerical calculations show that the integrals can be cut at $R_{min}=100~a_0$ and $R_{max}=500,~800,~1000~a_0$ for v=0,~1 and 2 (see Figures \ref{fig:Potpot} and \ref{fig:wf}a). Asymptoptic expansions for the ground state wave functions can therefore be used. Their accuracies were checked by comparing the results obtained in this way with the shift calculated using the wavefunctions of ab initio potentials \cite{Starckmeyer,Gadea2002,Gadea2004,Przybytek}. The calculation of the couplings $\Omega_{S,M_S;\ell,M_\ell}(r)$ is reported in appendix B.

\subsection{\label{sec:swave}Contributions of $s$-wave ground states}
In the experiment described in \cite{Kim2005} the excited state is probed by the excitation of atoms colliding in the $S=2,M_S=2;\ell=0,M_\ell=0$ channel with a scattering energy $E_\infty$ using $\sigma^-$ light. The contribution of this channel to the shift involves the regular and irregular solutions of the radial Schr\"odinger equation at energy $E=E_{\infty}=(k^\infty_{M_{S}=2})^2/2\mu$ for the $^5\Sigma_g^+$ potential and $\ell=0$. These solutions are shown in Figure \ref{fig:wf}b. In the experiment described in \cite{Kim2005}, the colliding energy is small enough for having the $s$-wave phase-shift $\eta_{S=2,M_S=2;\ell=0,M_\ell=0}$ determined by $a$ ($\eta_{S=2,M_S=2;\ell=0,M_\ell=0} \simeq- k^\infty_{M_S=2} a$) and $k^\infty_{M_S=2} r <<1$ for $R_{min}\leq r\leq R_{max}$.  We can thus approximate the regular and irregular wavefunctions for the atoms interacting in the $^5\Sigma_g^+$ potential by their asymptotic expansion :
\begin{eqnarray}
\Psi_{2,2;0,0}(r)&=&\sqrt{\frac{2 \mu}{\pi k^\infty_{M_S=2}}}\sin \left[ k^\infty_{M_S=2}(r-a)\right]\label{eq:sf} \\
\tilde{\Psi}_{2,2;0,0}(r)&=&-\sqrt{\frac{2 \mu}{\pi k^\infty_{M_S=2}}}\cos \left[ k^\infty_{M_S=2}(r-a)\right]
\label{eq:sg} \end{eqnarray}
Using equation (\ref{eq:MessiahShift}) and developing (\ref{eq:sf}) and (\ref{eq:sg}) for $k^\infty_{M_S=2} \rightarrow 0$ gives the following equations :
\begin{eqnarray}
\delta^{v=i}_{2,2;0,0}(E_{\infty})\approx -4\mu\int_0^\infty dr_1 \int_0^{r_1} dr_2  \Omega_{2,2;0,0}(r_1)\phi^{v=i}_e(r_1)\Omega_{2,2;0,0}(r_2)\phi^{v=i}_e(r_2) \nonumber \\
(r_2-a)\left[1+\frac{(k^\infty_{M_{S}=2})^2}{2}((r_1-a)^2+\frac{(r_2-a)^2}{6})\right] \label{eq:approxcontribs}
\end{eqnarray}
Equation (\ref{eq:approxcontribs}) shows that when $k^{\infty}_{M_S=2}\rightarrow 0$, the $s$-wave contribution to the shift contains terms linear in $a$ and quadratic in $k^\infty_{M_S=2}$ and thus linear in $E_{\infty}$. In the limit $k^{\infty}_{M_S=2}=0$, equation (\ref{eq:approxcontribs2}) shows explicitly the linear dependence of $\delta^{v=i}_{2,2;0,0}$ as a function of $a$:
\begin{eqnarray}
\delta^{v=i}_{2,2;0,0}(0)= \left[-4\mu\int_0^\infty dr_1 \int_0^{r_1} dr_2  \Omega_{2,2;0,0}(r_1)\phi^{v=i}_e(r_1)\Omega_{2,2;0,0}(r_2)\phi^{v=i}_e(r_2)r_2\right]-\nonumber\\
a\left[-4\mu\int_0^\infty dr_1 \int_0^{r_1} dr_2  \Omega_{2,2;0,0}(r_1)\phi^{v=i}_e(r_1)\Omega_{2,2;0,0}(r_2)\phi^{v=i}_e(r_2)\right]\label{eq:approxcontribs2}
\end{eqnarray}

The state $\ket{S=2,M_S=0;\ell=0,M_\ell=0}$ is also coupled by $\sigma^+$ light to the excited state. Its contribution to the shift involves wavefunctions which have the same form as (\ref{eq:sf}) and (\ref{eq:sg}), where $k^\infty_{M_S=2}$ has to be replaced by $k^\infty_{M_S=0}=\sqrt{2\mu(E_\infty+2g_S\mu_B B)}$ according to (\ref{eq:Einf}) and (\ref{eq:kinf}). In case of a large magnetic field, the approximation $\eta_{S=2,M_S=2;\ell=0,M_\ell=0} \simeq- k^\infty_{M_S=0} a$ and the inequality $k^\infty_{M_S=0} r <<1$ for $R_{min}\leq r\leq R_{max}$, may no longer be valid, and numerical calculations of the wavefunctions should be used to replace the formula (\ref{eq:approxcontribs}). The same remark holds for the state $\ket{S=0,M_S=0;\ell=0,M_\ell=0}$. Additionally, the phase shift in this channel is related to the scattering length $a_1$ which is estimated to be $20~a_0\leq a_1 \leq 50~a_0$ according to previous calculations \cite{Muller,Leocollision}. The uncertainty in $a_1$ induces an uncertainty on the calculation of the shift if $\sigma^+$ polarization is present in the PA laser.

\subsection{\label{sec:dwave}Contributions of $\ell \neq 0$ waves ground states}
The regular and irregular wavefunctions for a channel $\ket{S,M_S;\ell\neq 0,M_\ell}$ are shown in Figure (\ref{fig:wf}c). Although the amplitude of the regular wavefunction is negligible compared to the $s$-wave case, the one of the irregular wavefunction is not. This means that the contribution of an $\ell\neq 0$ channel to the shift, which involves the product of these wavefunctions (see equation (\ref{eq:MessiahShift})) may not be negligible compared to the $s$-wave case. The contributions of the $S,M_S;\ell\neq 0,M_\ell$ channels are estimated with a very good accuracy by neglecting the electronic interaction potential compared to the centrifugal potential when integrating the radial Schr\"odinger equation. This shows that these contributions are independent of the interaction potential and thus independent of the $s$-wave scattering length. As the range $R_{min}<r<R_{max}$ lies in the classically forbidden region in the centrifugal barrier, the exact phase shift in the $\ell>0$ wave does not need to be known. The regular and irregular wavefunctions are very well approximated by regular and irregular spherical Bessel functions $j_\ell(k^\infty_{M_S}r)$ and $y_\ell(k^\infty_{M_S}r)$. 

\begin{eqnarray}
\Psi_{S,M_S;\ell\neq 0,M_\ell}(r)&=&\sqrt{\frac{2\mu}{\pi k^\infty_{M_S}}}\left(k^\infty_{M_S}r\right) j_{l}(k^\infty_{M_S}r) \label{eq:dgf} \\
\tilde{\Psi}_{S,M_S;\ell\neq 0,M_\ell}(r)&=&\sqrt{\frac{2\mu}{\pi k^\infty_{M_S}}}\left(k^\infty_{M_S}r\right) y_{l}(k^\infty_{M_S}r)
\label{eq:dgg}
\end{eqnarray}

Using the expansion of the wavefunctions (\ref{eq:dgf}) (\ref{eq:dgg}) for small $k^\infty_{M_S}$, we find the following dependence on scattering energy and magnetic field :

\begin{eqnarray}
\fl\delta_{S,M_S;\ell\neq0,M_\ell}^{v=i}\approx -4\mu\int_0^\infty dr_1 \int_0^{r_1} dr_2 \Omega_{S,M_S;\ell\neq 0,M_\ell}(r_1)\Omega_{S,M_S;\ell\neq 0,M_\ell}(r_2)\phi^{v=i}_e(r_1)\phi^{v=i}_e(r_2) \nonumber \\
\frac{1}{(2\ell+1)}\frac{r_2^{\ell+1}}{r_1^\ell}(1+\frac{(k^{\infty}_{M_S})^2}{2}\frac{(2\ell-1)r_2^2-(2\ell+3)r_1^2}{(2\ell-1)(2\ell+3)}) \label{eq:approxcontribl}
\end{eqnarray}

\subsection{\label{sec:results}Results}
The results (\ref{eq:approxcontribs}) and (\ref{eq:approxcontribl}) for each channel $S,M_S;\ell,M_\ell$ are summed according to (\ref{eq:sumshift}) to give the final result. Table \ref{tab:contribsigmamin} gives the $a$ dependence of the shift in the case $E_\infty=0$, $B=0$ and $\sigma^-$ polarization of light. It shows the partial contributions of the $\ell=0,2,4$ collisions, their sum, and the total shift for $a=143~a_0$ \cite{Moal2006}. The $s$-wave contribution to the shift is dominant, and according to section \ref{sec:swave} it is also independent of the magnetic field. As a consequence, these results are quantitatively similar to what is obtained in the experimental conditions of the experiment described in \cite{Kim2005}.
\begin{table}[htbp]
\caption{\label{tab:contribsigmamin}
Contributions to the shift in $\mathrm{MHz/W.cm^{-2}}$ due to $s$,$d$, and $g$-wave collisional channels in case of coupling by $\sigma^{-}$ light, with a magnetic field $B=0$, and with a scattering energy $E_\infty=0$. $a$ is the scattering length for the ground-state $^5\Sigma_g^+$ potential expressed in unit of $a_0$. The last column gives the result for $a=143~a_0$ which is the value measured in \cite{Moal2006}.}
\begin{indented}
\item[]\begin{tabular}{cccccc}
 &$s$-wave&$d$-wave&$g$-wave&total&total~($a=143~a_0$) \\
\hline
v=0& -16.75+0.0862~$a$ & -1.89 & -0.06 & -18.70+0.0862~$a$ & -6.37 \\
v=1&-13.47+0.0310~$a$ &-2.64 & -0.02 & -16.13+0.0310~$a$ & -11.70  \\
v=2&-38.16+0.1001~$a$ &-5.81 & -0.01 & -43.88+0.1001~$a$ & -29.57
\end{tabular}
\end{indented}
\end{table}
\begin{figure}
\begin{center}
\includegraphics[scale=0.8]{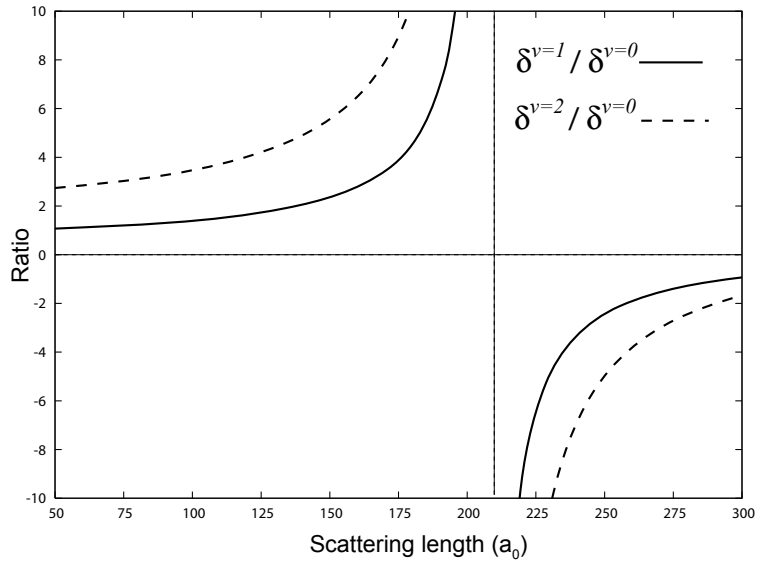}
\caption{\label{fig:ratio}Ratio of the shifts $\delta^{v=1}/\delta^{v=0}$ and $\delta^{v=2}/\delta^{v=0}$ as a function of the scattering length $a$ in case of a $\sigma^-$ coupling and with a magnetic field $B=0$. At $a\approx 210~a_0$, $\delta^{v=0}$ changes sign. For $a \leq 210~a_0$ the level $v=0$ is shifted towards red detuning which was also observed experimentally \cite{Kim2005}.}
\end{center}
\end{figure}

As mentioned earlier, the level shift is a linear function of the coupling elements $\Omega_{S,M_S;l,M_l}^2$, and so a linear function of the laser intensity $I$. However, the calibration of the laser intensity is difficult to achieve experimentally.  The ratio of the shifts for different levels is easier to measure accurately. Figure \ref{fig:ratio} shows the ratio of the shifts $\delta^{v=1}/\delta^{v=0}$ and $\delta^{v=2}/\delta^{v=0}$ as a function of the scattering length $a$ in case of the coupling with a purely $\sigma^-$ light and for a magnetic field $B=0$. The singularity at $a\simeq210~\mathrm{a_0}$ is associated to the cancellation and the change of sign of $\delta^{v=0}$. The curves of Figure \ref{fig:ratio} adapted to the $B\neq0$ case have been used in \cite{Kim2005} to deduce from these measurements a range of values for the scattering length $a$. The measured ratios in \cite{Kim2005} are  $\delta^{v=1}/\delta^{v=0}=1.70\pm0.14$ and $\delta^{v=2}/\delta^{v=0}=3.3\pm0.7$, and correspond to the left part of the figure where all $\delta^{v}$ are towards red detuning. Let us notice that Table \ref{tab:contribsigmamin} is given for $k^\infty_{M_S=2}=0$, so that the linear dependence of the $s$-wave contribution to the shift is linear in $a$ for any value of $a$. In the experiment described in \cite{Kim2005}, $k^\infty_{M_S=2}(r-$a$)<<1$ for $R_{min}<r<R_{max}$ and $a$ in the expected range for spin-polarized metastable helium. Therefore, according to section \ref{sec:swave}, the $s$-wave contribution to the shift can still be considered as a linear function of $a$.

Table \ref{tab:contribsigmaplus} shows the results obtained with $\sigma^+$ polarization and $B=0$ as a function of the scattering length $a$. As each ground state channel is coupled either by $\sigma^-$ or $\sigma^+$ light, the corresponding contributions to the shift can be weighted by the amount of laser intensity of each polarization and then added up to account for a given experimental situation where both polarizations are present. We remind that as the experiment described in \cite{Kim2005} uses spin polarized atoms which form a $J=2$,$M_J=2$ colliding pair, and since we are interested in a $J'=1,M_{J'}=1$ excited state, the PA laser must necessarily contains a $\sigma^-$ component in order to excite the atoms. Non $\sigma^-$ components of light do not excite the atoms, but induce an extra shift of the excited states.
\begin{table}[htbp]
\caption{\label{tab:contribsigmaplus}
Contributions to the shift in $\mathrm{MHz/W.cm^{-2}}$ due to $s$,$d$, and $g$-wave collisional channels in case of coupling by $\sigma^{+}$ light, with a magnetic field $B=0$, and with a scattering energy $E_\infty=0$. $a$ is the scattering length for the ground-state $^5\Sigma_g^+$ potential expressed in unit of $a_0$. The uncertainty in the $s$-wave contribution is due to the uncertainty on the scattering length of the ground state $^1\Sigma_g^+$ potential $a_1=35\pm15~a_0$(see section \ref{sec:swave}). The last column gives the results for $a_1=35~a_0$ and $a=143~a_0$ \cite{Moal2006}.}
\begin{indented}
\item[]\begin{tabular}{cccccc}
 &$s$-wave&$d$-wave&$g$-wave&total&total~($a=143~a_0$) \\
\hline
v=0& -5.86$\pm$0.03+0.0144 $a$ & -3.49 & -0.01 & -9.36$\pm$0.03+0.0144 $a$ & -7.36$\pm$ 0.03\\
v=1& -6.89$\pm$0.02+0.0052 $a$ & -4.10 & -0.004 & -10.99$\pm$0.02+0.0052 $a$ & -10.25$\pm$ 0.02 \\
v=2& -17.80$\pm$0.06+0.0167 $a$ & -8.70 & -0.002 & -26.50$\pm$0.06+0.0167 $a$ & -24.11$\pm$ 0.06
\end{tabular}
\end{indented}
\end{table}

\begin{figure}
\begin{center}
\includegraphics[scale=0.8]{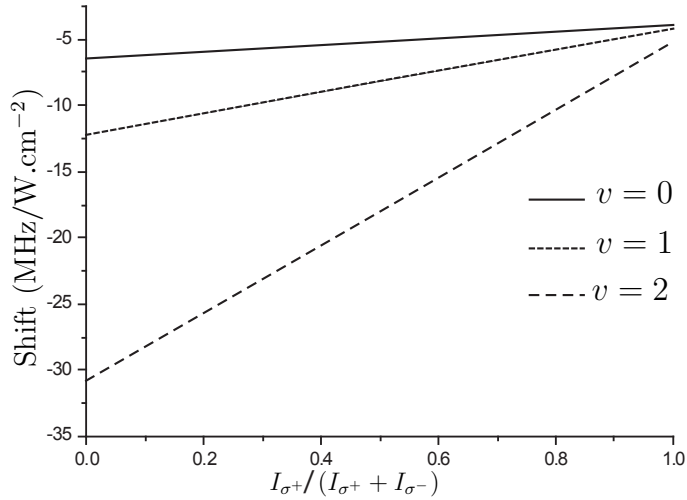}
\caption{\label{fig:polar} Influence of the polarization of light on the shift of $v=0,1,2$ for a magnetic field corresponding to a Zeeman shift of 20 MHz between the $M_S=2$ and $M_S=0$ channels, a scattering energy $E_\infty/h=0.1$~MHz (experimental conditions of the experiment described in \cite{Kim2005}) and a scattering length $a=143~a_0$ \cite{Moal2006}. $I_{\sigma^\pm}$ is the laser intensity in the $\sigma^\pm$ polarization. For $v=2$ the shift is more dependent on the polarization than for $v=0,1$. This is due to the fact that the value of the integral (\ref{eq:MessiahShift}) is approximately zero for $v=2$ and for the scattering wavefunction for the $\sigma^+$ coupled channel $S=2,M_S=0;\ell=0,M_\ell=0$ at energy $(k^\infty_{M_S=0})^2/2\mu$. Therefore the $s$-wave contribution to the shift is strongly reduced: $|\delta^{v=2}|$ is much smaller for a $\sigma^+$ polarization than for a $\sigma^-$ one.}  
\end{center}
\end{figure}

In zero magnetic field, we can notice that the results of the last column in Table \ref{tab:contribsigmaplus} are less than 15 percent different from those of Table \ref{tab:contribsigmamin}. However in the experimental situation of \cite{Kim2005}, the magnetic field is non zero, and $k^\infty_{M_S=0}R_{max}\simeq 1$  so that the approximation mentioned in section \ref{sec:swave} breaks down. Numerical calculations of the scattering wavefunctions show that the $s$-wave contribution to the shift is strongly suppressed  for $v=2$ with $\sigma^+$ polarization, and the overall shift is significantly reduced in absolute value compared to the $\sigma^-$ coupling case. This is illustrated in Figure \ref{fig:polar}. The shifts for $v=0,1,2$ are plotted versus the polarization of light in the conditions of the experiment described in \cite{Kim2005}, where linear polarization was investigated in addition to $\sigma^-$ polarization. 
 
\subsection{Additional light shifts}
All the previous calculations deal with the light-induced frequency shift of the vibrational level which is excited by the PA laser in the $0_u^+$ potential. The initial scattering state in the $^5\Sigma_g^+$ potential is also shifted by light. When the two colliding atoms are far apart, the non resonant laser excitation, detuned by an amount $\Delta<0$ from the atomic $S-P_0$ transition, induces a shift of the internal atomic energies which is of the order of $\Omega_\infty^2/\Delta$, where $\Omega_\infty$ is the Rabi coupling at large distance. For the photoassociated bound level $v=0,1,2$ $|\Delta|>200~\mathrm{MHz}$ is much larger than the width $\gamma/2\pi=1.62~\mathrm{MHz}$ of the excited atomic $P$ state, so that $\gamma$ does not appear in the expression of the shift. Since $\Delta$ is negative, the shift is negative and produces an additional blue shift of the photoassociation line. This shift is found to be less than 20 percent of the shift calculated in section \ref{sec:results}. It depends on the photoassociated bound-level as it depends on $\Delta$. We calculated the quantity $\delta^v/(\Omega_\infty^2/\Delta^v)$ in the conditions of the experiment described in \cite{Kim2005}, where $\delta^v$ is the shift of the level $v$ calculated in section \ref{sec:calc}, $\Delta^v$ is the laser detuning associated to the excited bound level $v$. This quantity has a very small dependence on the bound level $v$. As a consequence, the ratio of the shifts corresponding to different vibrational levels, which is used to determine the scattering length (see Figure \ref{fig:ratio}) is calculated to be the same within 3 percent, whether this additional shift is taken into account or not. The difference is well below the experimental accuracy of the measurements described in \cite{Kim2005} and we have neglected it.

A precise treatment of this effect would require a description of the PA as a scattering problem, and to relate the experimental signal to the loss of atoms, i.e to the loss of unitarity of the $S$-matrix. ($1-\sum_{S',M_S',\ell',M_\ell'}\left|S_{S',M_S',\ell',M_\ell';S,M_S;\ell,M_\ell} \right|^2$) \cite{Napolitano1994}. This loss rate exhibits a resonance which is shifted from its position in the zero intensity limit by the quantity $\delta$ from equations (\ref{eq:sumshift}) and (\ref{eq:MessiahShift}), whatever the incoming channel $S,M_S;\ell,M_\ell$ is \cite{Simoni}. The difficulty which is then encountered is that the scattering channels remain coupled at infinite atom separation due to the non resonant excitation of the atomic $S-P_0$ transition. A change of basis must therefore be introduced in order to use states which are uncoupled at infinite separation (dressed states), as they should be in a $S$-matrix description. Some specific features linked to the proper definition of the boundary conditions in photoassociation called \emph{dressing effects} were investigated in \cite{Julienne1988,Napolitanodressed,Naposolo,Montalvao}. However, photoassociative frequency shifts were not studied in this context. Treating them would require to take into account more excited channels, for example the $1_u$ repulsive potential connected to the $S-P_0$ asymptote. It is involved in the non-resonant laser excitations at large distance since it is degenerate with the purely long range $0_u^+$ potential in the limit of infinite interatomic distance. The present accuracy of the experiment does not yet require a more precise calculation of this type.


\section{Conclusion}
In this article we have given a detailed theoretical analysis of the light-induced frequency shifts in the PA of $^4He^*$ atoms. We have considered the case  where a pair of spin polarized  atoms are excited into a molecular state in the purely long range $J=1,0_u^+$ potential by absorbing one photon at a frequency red to the $2^3S_1-2^3P_0$ transition. The coupling of this excited state to ground bound levels and scattering states has been considered and the physical role played by theses various contributions has been discussed. We have also emphasized the importance of the contribution of the $\ell\neq0$ scattering channels. Explicit expressions of the light shifts have been derived from an effective Hamiltonian approach. Results have been given for several vibrational levels ($v=0,1,2$) in the $J=1,0_u^+$ potential and for both  $\sigma^-$ polarization of the laser light, which is actually absorbed by the spin polarized atoms, and $\sigma^+$ polarization, which is not absorbed but contributes to the shift through the coupling to various channels. The explicit calculation of the shift ratios for two different values of v, which are independent of the laser intensity, provided an appropriate tool to interpret previous experimental results of our group and allowed us to derive a value for the scattering length $a$\cite{Kim2005}. 

Actually the $^4He^*$ atom can be regarded as a model system for simply solving the problem of the light shifts in a photoassociation experiment. First, the entrance channel for the pair of incoming atoms is well defined, as the atoms are spin polarized and the temperature is ultra low. Second, the helium atom has no hyperfine structure, which limits the number of coupling channels to be considered. In addition, the fact that the excited molecular state $0_u^+$ is purely long range allows an exact calculation of the excited molecular potential and wave functions, and the use of an asymptotic expansion of the wave function for the incoming pair of ground state atoms thus avoiding its complexity at short interatomic distance. This is why one simply finds a linear dependence of the shifts as a function of the $s$-wave scattering length $a$. Measuring the light-induced frequency shifts is thus a very appropriate method to deduce a value for $a$ without having to precisely know the ground state potential. In that respect the light-shift method is more direct yet much less precise than the two-photon Raman spectroscopy method \cite{Moal2006}, in which the relation between the measured energy of the least bound state and the value of $a$ has to be infered from the ground state potential.

Alkaline-earth atoms are other model systems in which light shifts could be studied. They are currently the focus of intense studies. They appear as simpler systems than alkalis since they have isotopes lacking hyperfine structure. From light shifts one could infer complementary or more accurate results on scattering lengths than those given by other methods. One could also test photoassociation theories on these simple systems. This could for instance apply to $Ca$, $Sr$ or $Yb$ atoms. The isotopes $^{40}Ca$, $^{86}Sr$, $^{88}Sr$ and $^{174}Yb$ have been studied in \cite{Degenhardt2003,Mickelson2005,Yasuda2006, Takasu2004,Tojo2006}. The case of $^{88}Sr$ is of special interest as it has long range potentials which were investigated in \cite{Nagel2005}, and the case of $^{174}Yb$ is specific as it displays atomic levels without fine nor hyperfine structures.  

\ack
The authors would like to thank Paul Julienne and Yvan Castin for interesting discussions, and Christian Buggle for careful reading of the manuscript.

\appendix
\section{\label{sec:appsym} Symmetrized states for collisions of metastable atoms}
The physical states to be considered for describing the collisions between $^4He^*$ atoms have to be symmetric under the permutation of the bosonic identical atoms. These are obtained using the symmetrization operator $(1+P_{ab})/2$ acting on the non symmetrized basis (\ref{eq:SMSlMlbasis}). Using the transformation in the spin space from the $\ket{(S_a,S_b)S,M_S}$ to the $\ket{S_a,M_{S_a};S_b,M_{S_b}}$ basis, where $S_{a}$,$S_{b}$ and $M_{S_{a}}$,$M_{S_{b}}$ are the spins of atoms $a$, $b$ and their projection on the lab-fixed axis respectively. Using also a Clebsch-Gordan identity, we get :
\begin{eqnarray*}
\ket{(S_a,S_b)S,M_S}&=&\sum_{M_{S_a},M_{S_b}}\braket{S_a,M_{S_a};S_b,M_{S_b}}{(S_a,S_b)S,M_S}\ket{S_a,M_{S_a};S_b,M_{S_b}} \\
~&=&\sum_{M_{S_a},M_{S_b}} (-1)^{S_a+S_b-S}\braket{S_b,M_{S_b};S_a,M_{S_a}}{(S_b,S_a)S,M_S} \\
~&&\ket{S_a,M_{S_a};S_b,M_{S_b}}
\end{eqnarray*}
We deduce the action of $P_{ab}$ on the spin part:
\begin{equation*}
P_{ab}\ket{(S_a,S_b)S,M_S}=(-1)^{S_a+S_b-S}\ket{(S_b,S_a)S,M_S}
\end{equation*}
The action of $P_{ab}$ on the spatial part is given by:
$$P_{ab}\ket{\ell,M_\ell}=(-1)^\ell\ket{\ell,M_\ell}$$
We get the properly symmetrized basis :
\begin{eqnarray}
\fl\frac{1+P_{ab}}{2}\ket{(S_a, S_b)S, M_S;\ell,M_\ell}=\frac{1}{2}\ket{(S_a,S_b)S, M_S;l,M_l}+\frac{(-1)^{S_a+S_b-S+\ell}}{2}\ket{(S_b, S_a)S, M_S;\ell,M_\ell} \nonumber \\
~ 
\label{eq:Slbasis}
\end{eqnarray}
In the case of $^4He^*$, $S_a=S_b=1$, which implies $S=0,1,2$ (singlet, triplet and quintet states). We will therefore use a simplified notation:
\begin{equation}
\fl\ket{S, M_S;\ell,M_\ell}=\frac{1}{2}\ket{(S_a,S_b)S, M_S;l,M_l}+\frac{(-1)^{-S+\ell}}{2}\ket{(S_b, S_a)S, M_S;\ell,M_\ell} 
\label{eq:Slbasis2}
\end{equation}
The formula (\ref{eq:Slbasis2}) also shows that singlet ($S=0$) and quintet ($S=2$) states collide in even $\ell$ partial waves, and triplet states in odd $\ell$ partial waves.

We now investigate the properties of the collisional state under the inversion of the electrons coordinates through the center of charge of the molecule (gerade/ungerade symmetry). In the case of homonuclear dimers, the center of charge and the center of mass are the same. As both the $^4He^*$ atom and its nucleus are bosons, the exchange of the atoms and the exchange of the nuclei of the quasi-molecule should not change the wavefunction of the pair of atoms. The product of both operations is equivalent to the exchange of the electronic clouds and should therefore not change the total wavefunction. 

In the space coordinates the exchange of the electronic clouds is the product of the inversion of the electronic coordinates through the center of mass of the corresponding atom, and the inversion of the electronic coordinates through the center of mass of the molecule (gerade/ungerade symmetry). As the atoms are in a $2^3S_1$ state with an even spatial electronic wavefunction, the first operation is the identity. The second operation multiplies the wavefunction by $(-1)^w$, where $w=0$ for a gerade state, and $w=1$ for an ungerade state. 

In the spin space, the exchange of the electronic clouds multiplies the wavefunction by $(-1)^{S_a+S_b-S}=(-1)^S$.

As a consequence, $(-1)^{w+S}=+1$ due to the bosonic nature of the $^4He^*$ atom and of its nucleus. Singlet ($S=0$) and quintet ($S=2$) states are gerade states ($w=0$) whereas triplet states ($S=1$) are ungerade states ($w=1$)

\section{\label{sec:couplings}Light-induced couplings}
In this appendix we calculate the radiative coupling operators $\Omega_{eg}(r)$ introduced in section \ref{sec:formulas} and connecting the excited channel $e:(0_u^+;J'=1,M_{J'}=1)$ to the ground state channels $g:(S,M_S;\ell,M_\ell)$ given in equation (\ref{eq:coupledminusstates}) for a $\sigma^-$ excitation and in equation (\ref{eq:coupledplusstates}) for a $\sigma^+$ excitation. More precisely, the $\Omega_{eg}(r)$ are obtained by taking the matrix elements of the laser interaction Hamiltonian in the electric dipole approximation $-\sqrt{2\pi I/c}~\mathbf{d}.\boldsymbol{\epsilon}$ between \bra{e} and \ket{g}.
\begin{equation}
\fl\Omega_{eg}(r)=\Omega_{S,M_S;\ell,M_\ell}(r)=-\sqrt{2\pi I/c}\bra{J'=1,M_{J'}=1;0_u^+}\mathbf{d}.\boldsymbol{\epsilon}\ket{S,M_S;\ell,M_\ell}
\label{eq:eqom}
\end{equation}
In (\ref{eq:eqom}), $I$ is the laser intensity, $c$ is the speed of light, $\mathbf{d}$ the dipole operator of the quasi-molecule, $\boldsymbol{\epsilon}$ is the polarization vector of light. Note that the matrix element (\ref{eq:eqom}) is taken over the electronic variables and the angular variables of the molecular axis, so that $\Omega_{S,M_S;\ell,M_\ell}(r)$ remains an operator for the one-dimensional variable $r$, the distance between the two nuclei. Using the decomposition (\ref{eq:zerouj}) of the excited state, $\Omega_{S,M_S;\ell,M_\ell}(r)$ can be expressed as :

\begin{equation}
\fl\Omega_{S,M_S;l,M_l}(r)=-\sqrt{2\pi I/c}\sum_{i=1}^4 \alpha_{i}(r)\bra{J'=1,M_{J'}=1,i} \mathbf{d}.\boldsymbol{\epsilon}\ket{S,M_S;l,M_l}
\label{eq:finalcoupling}
\end{equation}
In equation (\ref{eq:finalcoupling}), the index $i$ stands for one of the electronic states $^5\Sigma_g^+$,$^5\Pi_u$,$^3\Pi_u$,$^1\Sigma_u^+$ appearing in the expansion of the excited molecular state \ket{J'=1,M_{J'}=1,0_u^+}, the $\alpha_i$ coefficients are found by the diagonalization of the electronic Hamiltonian for a $S+P$ molecule (see subsection \ref{sec:excited}). The $\bra{J'=1,M_{J'}=1,i} \mathbf{d}.\boldsymbol{\epsilon}\ket{S,M_S;l,M_l}$ matrix elements are needed to calculate (\ref{eq:finalcoupling}).

The electric dipole operator acts only on the orbital angular momentum of the electrons, and can therefore be conveniently expressed in the Hund's case (a) basis for the rotating molecule (\ref{eq:casa}), in which its projection on the molecular axis is well defined.
\begin{equation}
\ket{J,M_J,\Omega,S,\Sigma,\Lambda,w}\equiv\sqrt{\frac{2J+1}{4\pi}}D^{J~*}_{M_J,\Omega}(\alpha,\beta,\gamma=0)\ket{S,\Sigma,\Lambda,w}
\label{eq:casa}
\end{equation}
$\vec{J}=\vec{L}+\vec{S}+\vec{\ell}$ is the total angular momentum, which is the sum of the orbital angular momentum of the electrons $\vec{L}$, the spin of the electrons $\vec{S}$ and the angular momentum for the rotation of the atoms $\vec{\ell}$ which has a zero projection on the molecular axis. $M_J$ is the projection of $\vec{J}$ on the lab-fixed quantization axis, $\Lambda$ and $\Sigma$ the projection of $\vec{L}$ and $\vec{S}$ on the molecular axis, $\Omega=\Lambda+\Sigma$ is the projection of $\vec{J}$ on the molecular axis. The transformation of the wavefunction under the inversion of the electronic space coordinates is characterized by a quantum number $w$ such that $(-1)^w=1$ (gerade symmetry) or $(-1)^w=-1$ (ungerade symmetry). $D^{J~*}_{M_J,\Omega}$  is a matrix element of the $(2J+1)\times(2J+1)$ unitary rotation matrix which is a function of the Euler angles $\alpha$,$\beta$,$\gamma$ and which defines the relationship of the space-fixed lab frame with that of the body-fixed molecule frame. $\gamma=0$ is a convention for linear molecules for which two angles are enough to describe the rotation.

The states wich are involved in the decomposition (\ref{eq:finalcoupling}) are related to the basis (\ref{eq:casa}) by the relations (\ref{eq:touz2casa}):
\begin{eqnarray}
\ket{J'=1,M_{J'}=1,^5\Sigma_u^+}&:&\ket{1,1,0,2,0,0,1} \nonumber\\
\ket{J'=1,M_{J'}=1,^1\Sigma_u^+}&:&\ket{1,1,0,0,0,0,1} \nonumber\\
\ket{J'=1,M_{J'}=1,^3\Pi_u}&:&1/\sqrt{2}\left(\ket{1,1,0,1,1,-1,1}-\ket{1,1,0,1,-1,1,1} \right) \nonumber\\
\ket{J'=1,M_{J'}=1,^5\Pi_u}&:&1/\sqrt{2}\left(\ket{1,1,0,2,1,-1,1}+\ket{1,1,0,2,-1,1,1} \right) \label{eq:touz2casa}
\end{eqnarray}
In (\ref{eq:touz2casa}), the linear superposition for the $\Pi_u$ states is formed so that the electronic states have (+) symmetry under the reflexion of the space and spin electronic coordinates about a plane containing the molecular axis, and are thus part of the $0_u^+$ subspace.

The dipole operator $\mathbf{d}$ is best expressed in the body-fixed molecular frame, and $\boldsymbol{\epsilon}$ in the space-fixed frame. The product $\mathbf{d}. \boldsymbol{\epsilon}$ can be expressed using a rotation matrix $D$, using the quantum numbers $n$ and $m$ which refer to the projection of $\mathbf{d}$ on the molecular basis and $\boldsymbol{\epsilon}$ on the space fixed basis ($m=0,+1,-1$ for $\pi$,$\sigma^+$,$\sigma^-$ polarization) \cite{Burke}.
\begin{equation}
\mathbf{d}.\boldsymbol{\epsilon}=\sum_{m,n=0,\pm 1}(-1)^m\epsilon_{-m}d_n D^{1~*}_{m~n}(\alpha,\beta,0)
\label{eq:de}
\end{equation}
The projections in the molecular frame $d_{n=0,\pm1}$ of the operator $\mathbf{d}$ act on the electronic part $\ket{S,\Sigma,\Lambda=0,w}$ of a Hund's case (a) ground state according to equation (\ref{eq:dipphi}), where $d_{at}$ is the atomic electric dipole moment for the $S-P$ transition \cite{Burke}.
\begin{equation}
\bra{S',\Sigma',\Lambda',w'}d_n$\ket{S,\Sigma,\Lambda=0,w}$=\frac{1+(-1)^{w'-w+1}}{\sqrt{2}}d_{at}\delta_{S,S'}\delta_{\Sigma,\Sigma'}\delta_{\Lambda',n}
\label{eq:dipphi}
\end{equation}
Using (\ref{eq:casa}),(\ref{eq:de}),(\ref{eq:dipphi}), we can form the expression of the action of the electric dipole operator in the Hund's case (a) basis for the rotating molecule.
\begin{eqnarray*}
\fl\bra{J',M_J',\Omega',\Lambda',S',\Sigma',w'}\mathbf{d}.\boldsymbol{\epsilon}\ket{J,M_J,\Omega,\Lambda=0,S,\Sigma,w}=
\sqrt{\frac{(2J+1)(2J'+1)}{16\pi^2}}\\
\frac{1+(-1)^{w'-w+1}}{\sqrt{2}}d_{at}\delta_{S,S'}\delta_{\Sigma,\Sigma'}\delta_{\Lambda',n}\sum_{m,n}\epsilon_{-m}(-1)^m  \\
\left[\int_0^{2\pi}d\alpha\int_0^{\pi}\sin{\beta}d\beta D^{J'}_{M_J'~\Omega'}(\alpha,\beta,0)D^{1~*}_{m~n}(\alpha,\beta,0)D^{J~*}_{M_J~\Omega}(\alpha,\beta,0)\right]
\end{eqnarray*}
The evaluation of the integral over rotation matrix elements gives \cite{Burke}:
\begin{eqnarray}
\bra{J',M_J',\Omega',\Lambda',S',\Sigma',w'}\mathbf{d}.\boldsymbol{\epsilon}\ket{J,M_J,\Omega,\Lambda=0,S,\Sigma,w}=
\sqrt{\frac{2J+1}{2J'+1}}\nonumber\\
\sum_{m,n}\epsilon_{-m}(-1)^m \braket{J,M_J;1,m}{J',M_{J'}}\braket{J,\Omega;1,n}{J',\Omega'}\nonumber\\
\frac{1+(-1)^{w'-w+1}}{\sqrt{2}}d_{at}\delta_{S,S'}\delta_{\Sigma,\Sigma'}\delta_{\Lambda',n}
\label{eq:decasa}
\end{eqnarray}



Now, the transformation of the electronic ground state between the basis \ket{S,M_S;\ell,M_\ell} (\ref{eq:Slbasis2}) and the Hund's case (a) basis (\ref{eq:casa}) with $\Lambda=0$ is needed. The rotation in the spin states between the lab-fixed frame and the molecular frame is expressed in the relation (\ref{eq:rotSMS}).
\begin{equation}
\ket{S;M_S}=\sum_\Sigma D^{S*}_{M_S,\Sigma}(\alpha,\beta,0)\ket{S,\Sigma}
\label{eq:rotSMS} 
\end{equation}
A state \ket{\ell,M_\ell} is associated to the spherical harmomic $Y_{\ell,M_\ell}$ which can be expressed using a rotation matrix element (Appendix C in \cite{Messiah}):
\begin{equation}
\ket{\ell;M_\ell}\leftrightarrow Y_{\ell,M_\ell}(\beta,\alpha)=\sqrt{\frac{2\ell+1}{4\pi}}D^{\ell *}_{M_\ell,0}(\alpha,\beta,0)
\label{eq:lMl}
\end{equation} 
In the electronic ground state, the orbital angular momentum of the electrons is zero so that $\Lambda=0$. The property under gerade symmetry is given by $(-1)^{w+S}=+1$ (see appendix A). We write therefore using (\ref{eq:rotSMS}) and (\ref{eq:lMl}):
\begin{equation}
\fl \ket{S,M_S;\ell,M_\ell}\equiv\sqrt{\frac{2\ell+1}{4\pi}}D^{\ell *}_{M_\ell,0}(\alpha,\beta,0)\\
\sum_\Sigma D^{S*}_{M_S,\Sigma}(\alpha,\beta,0)\ket{S,\Sigma,\Lambda=0,w=S}
\end{equation}

The following scalar product is deduced.
\begin{eqnarray}
\fl\braket{J,M_J,\Omega,S',\Sigma,\Lambda=0,w}{S,M_S;\ell,M_\ell}=\sqrt{\frac{(2J+1)(2\ell+1)}{16\pi^2}}\frac{(1+(-1)^{w+S})}{2}\delta_{S,S'} \nonumber \\ \left[\int_0^{2\pi}d\alpha\int_0^{\pi}\sin{\beta}d\beta D^{J}_{M_J,\Omega}(\alpha,\beta,0)D^{\ell~*}_{M_\ell,0}(\alpha,\beta,0)D^{S~*}_{M_S,\Sigma}(\alpha,\beta,0)\right]  
\end{eqnarray}
Performing the integration of the product of rotation matrix elements, we obtain the transformation (\ref{eq:SScasa}) between the basis (\ref{eq:Slbasis2}) and (\ref{eq:casa}):
\begin{eqnarray}
\fl \braket{J,M_J,\Omega,S',\Sigma,\Lambda=0,w}{S,M_S;\ell,M_\ell}=\sqrt{\frac{2\ell+1}{2J+1}} \delta_{S,S'}\delta_{w,S} \braket{S,M_S;\ell,M_\ell}{J,M_J}\braket{S,\Sigma;\ell,0}{J,\Omega}  \nonumber \\
\label{eq:SScasa}
\end{eqnarray}

Using (\ref{eq:SScasa}) and (\ref{eq:decasa}), it is now possible to calculate the action of the electric dipole operator on the \ket{S,M_S;\ell;M_\ell} projected in the Hund's case (a) basis (\ref{eq:casa})
\begin{eqnarray*}
\fl\bra{J',M_J',\Omega',\Lambda',S',\Sigma',w'}\hat\mathbf{{d}}.\boldsymbol{\epsilon}\ket{S,M_S;\ell,M_\ell}=\sum_{m,n}\sum_{J,M_J,\Omega,\Sigma}(-1)^m\epsilon_{-m}\sqrt{2}d_{at} \frac{1+(-1)^{w'+S-1}}{2}\\
\delta_{\Lambda',n}\delta_{S,S'}\delta_{\Sigma,\Sigma'}\sqrt{\frac{2\ell+1}{2J'+1}}\braket{J,M_J;1,m}{J',M_J'}\braket{J,\Omega;1,n}{J',\Omega'}\braket{S,M_S;\ell M_\ell}{J,M_J}\\ \braket{S,\Sigma;l,0}{J,\Omega} 
\end{eqnarray*}
The summation can be simplified using the following remarks:
\begin{itemize}
\item The fourth Clebsch-Gordan coefficient is non zero for $\Omega=\Sigma$. Because of the coefficient $\delta_{\Sigma,\Sigma'}$, we have $\Omega=\Sigma'$
\item The third Clebsch-Gordan coefficient is non zero for $M_J=M_S+M_\ell$
\item The second Clebsch-Gordan coefficient is non zero for $n=\Omega'-\Omega=\Omega'-\Sigma'=\Lambda'$
\item The first Clebsch-Gordan coefficient is non zero for $m=M_{J'}-M_J=M_{J'}-M_S-M_\ell$
\end{itemize}
This gives finally equation (\ref{eq:finalde})
\begin{eqnarray}
\fl\bra{J',M_J',\Omega',\Lambda',S',\Sigma',w'}\hat\mathbf{{d}}.\boldsymbol{\epsilon}\ket{S,M_S;\ell,M_\ell}=\sum_{J}(-1)^{M_{J'}-M_S-M_\ell}\epsilon_{-M_{J'}+M_S+M_\ell}\sqrt{2}d_{at} \nonumber \\ \left( \frac{1+(-1)^{w'+S-1}}{2}\right)\sqrt{\frac{2\ell+1}{2J'+1}}\braket{J,M_S+M_\ell;1,M_{J'}-M_S-M_\ell}{J',M_{J'}}\nonumber \\ \braket{J,\Sigma';1,\Lambda'}{J',\Omega'}\braket{S,M_S;\ell M_\ell}{J,M_S+M_\ell}\braket{S,\Sigma';l,0}{J,\Sigma'}
\label{eq:finalde}
\end{eqnarray}

As a conclusion, using (\ref{eq:finalde}), and the relations (\ref{eq:touz2casa}), it is possible to calculate (\ref{eq:finalcoupling}). The \ket{J'=1,M_{J'}=1,^3\Pi_u} state does not contribute to the sum $(\ref{eq:finalcoupling})$ due to the selection rule on the $u/g$ symmetry (\ref{eq:dipphi}) and the fact that ground electronic states are gerade states (see \ref{sec:appsym}).

\bigskip

\bibliography{JPhysB}
\end{document}